\documentclass[aps,nofootinbib,prd,superscriptaddress,tightenlines,notitlepage,twocolumn,showpacs,floatfix]{revtex4-1}

\usepackage{dcolumn}
\usepackage{bm}
\usepackage{float}
\usepackage[colorlinks]{hyperref}
\usepackage{tabularx}
\usepackage{amssymb,amsmath,bm,tensor,braket}
\usepackage{xcolor}
\usepackage[varg]{txfonts}
\usepackage{enumerate}
\usepackage{mathtools}
\usepackage[capitalize]{cleveref}
\usepackage[utf8]{inputenc}
\usepackage[normalem]{ulem}
\usepackage[T1]{fontenc}
\usepackage[version=4]{mhchem}
\usepackage{natbib}
\usepackage{extarrows}
\usepackage{txfonts,newtxtext}
\usepackage{orcidlink}

\newcommand{\dd}{\mathrm{d}}

\begin{document}

\title{The $w$-mode instability of strangeon ergostars and its impact on energy extraction}

\author{Haojia Xia\,\orcidlink{0009-0004-9673-7062}}
\affiliation{Department of Astronomy, School of Physics, Peking University, Beijing 100871, China}%

\author{Hong-Bo Li\orcidlink{0000-0002-4850-8351}}
\affiliation{Lanzhou Center for Theoretical Physics, Key Laboratory of Theoretical Physics of Gansu Province, Key Laboratory of Quantum Theory and Applications of MoE, Gansu Provincial Research Center for Basic Disciplines of Quantum Physics, Lanzhou 730000, China}
\affiliation{Institute of Theoretical Physics \& Research Center of Gravitation, School of Physical Science and Technology, Lanzhou University, Lanzhou 730000, China}%

\author{Enping Zhou\,\orcidlink{0000-0002-9624-3749}}
\email[Corresponding author: ]{ezhou@hust.edu.cn}
\affiliation{School of Physics, Huazhong University of Science and Technology, Wuhan 430074, China}%

\author{Ren-Xin Xu\,\orcidlink{0000-0002-9042-3044}}
\email[Corresponding author: ]{r.x.xu@pku.edu.cn}
\affiliation{State Key Laboratory of Nuclear Physics and Technology, Peking University, Beijing 100871, China}%
\affiliation{Department of Astronomy, School of Physics, Peking University, Beijing 100871, China}%
\affiliation{Kavli Institute for Astronomy and Astrophysics, Peking University, Beijing 100871, China}%

\date{November 11, 2026}

\begin{abstract}
Rapidly rotating compact ergostars have been proposed as alternative central engines for short gamma-ray bursts. However, competing energy-loss channels, such as gravitational-wave emission driven by non-axisymmetric instabilities, could curtail the power available for relativistic electromagnetic outflows. In this work, we investigate the $w$-mode instability of rapidly rotating strangeon ergostars by solving the coupled spacetime and fluid perturbation equations in the Hilbert gauge using a spectral method, without relying on a slow-rotation approximation. We calculate the complex eigenvalues of the axial-led and polar-led $w$-mode branches originating from the fundamental $l=m=2$ modes of nonrotating stars along the maximum-energy-extraction paths of twelve representative strangeon equations of state. We find that both branches remain damped throughout these paths, despite the presence of ergoregions. We further vary the stellar angular velocity to determine the instability thresholds. In every case with an identified threshold, the angular velocity at the corresponding extraction-path endpoint remains below the critical value, with the largest ratio being $0.926$. Our results indicate that the $w$-mode instabilities examined here do not provide a growing gravitational-wave emission channel along the considered energy-extraction paths, supporting the availability of the previously estimated rotational energy reservoirs for electromagnetic extraction.
\end{abstract}

\maketitle

\section{Introduction}
\label{sec:intro}

The central engine of short gamma-ray bursts (sGRBs) after binary neutron star (BNS) mergers remains an open question in multi-messenger astrophysics \cite{GW170817,Savchenko_2017,2017GCN,Goldstein_2017}. Beyond the scenario of a nascent black hole enveloped by a hyperaccreting torus \cite{Blandford1977}, a rapidly rotating, horizonless compact star may serve as an alternative central engine \cite{Zhang_2001,Lü_2015}. A star possessing an ergoregion, where the stationary Killing vector becomes spacelike, is referred to as an ergostar. The existence of negative-energy states with respect to infinity may allow rotational energy to be extracted through processes analogous to those around rotating black holes \cite{Penrose2002,Komissarov2004,Komissarov2005}. Some earlier polytropic ergostar configurations were found dynamically unstable \cite{KEH2,Tsokaros2019}, while some recent numerical studies demonstrated the existence of dynamically stable ergostars and identified a high density near the stellar surface as a favorable condition for their formation \cite{Tsokaros2019,Tsokaros2020b}. These results motivate the investigation of self-bound compact stars as potential ergostar candidates.

A physically well-motivated realization of such self-bound matter is provided by strangeon stars (SnSs), in which quarks are conjectured to condense into localized, nucleon-like clusters (strangeons) mediated by non-perturbative strong interactions \cite{Xu:2003xe,Xu:2016uod,Lai:2017ney,2023AdPhX...837433L,2025IJMPA..4050180X}. Their phenomenological Lennard--Jones equation of state (EOS) can be sufficiently stiff to support massive, rapidly rotating configurations and features a finite density at the stellar surface \cite{Lai2009,Gao_2022,2022MNRAS.516.6172L,Yuan2025}. In our previous study \cite{Xia2026}, we systematically investigated uniformly rotating SnSs and identified an extensive region of ergostar configurations on the stable side of the turning-point boundary \cite{Stergioulas_1995,Friedman1988}. In the description adopted in that study, rotational energy extraction proceeds along a sequence of equilibrium configurations at constant baryon mass. As the star loses angular momentum, it moves along this sequence, which we refer to as an energy-extraction path. The path is followed only while the star retains an ergoregion and remains on the stable side of the turning-point boundary. By examining those possible evolutionary paths, we further estimated a potentially extractable rotational-energy reservoir of order $0.01\,M_\odot c^2$ ($\sim 10^{52}\,\mathrm{erg}$), sufficient to meet the energy requirements of typical sGRBs.

The existence of a substantial rotational-energy reservoir does not guarantee that this energy can power an sGRB. In particular, non-axisymmetric instabilities may provide competing channels for extracting rotational energy through gravitational-wave emission \cite{Chandrasekhar1970,FriedmanSchutz1978b}, thereby reducing the energy available for electromagnetic outflows. Of particular interest are spacetime-dominated $w$-modes, whose instability is closely associated with the presence of an ergoregion \cite{Friedman1978,Kokkotas_2004}. If such instabilities develop on timescales comparable to or shorter than that of electromagnetic energy extraction, they could significantly affect the energy budget of the central engine. It is therefore important to examine whether $w$-mode instabilities can arise along the proposed energy-extraction paths of strangeon ergostars.

The study of stellar $w$-modes was initiated by Kokkotas and Schutz, who identified a family of strongly damped spacetime oscillations in nonrotating relativistic stars~\cite{KokkotasSchutz1992}. Subsequent work established the existence of axial $w$-modes and refined the computation of complex frequencies for polar $w$-modes, particularly through numerical integration along complex radial contours to impose the outgoing-wave boundary condition~\cite{Kokkotas1994,Andersson1995}. Investigations of ultracompact stars further revealed long-lived trapped modes in both axial and polar spectra~\cite{AnderssonKojimaKokkotas1996}. The extension to rotating stars was formulated by Kojima, who derived the relativistic perturbation equations to first order in the stellar angular velocity~\cite{Kojima1992}. Early slow-rotation calculations also examined the rotational splitting of axial $w$-modes and the corresponding changes in their damping times~\cite{Andersson_1998}. These equations were later reformulated into systems suitable for numerical time evolution~\cite{Ruoff_2002eq}. Within the slow-rotation approximation, Kokkotas et~al.\ computed the complex frequencies of axial $w$-modes for uniform-density ultracompact stars and demonstrated that these modes can become unstable after the formation of an ergoregion, with growth times ranging from seconds to minutes~\cite{Kokkotas_2004}. Kr\"uger and Kokkotas derived the perturbation equations for rapidly rotating relativistic stars in the Hilbert gauge and developed a time-domain numerical scheme that includes dynamical spacetime perturbations. Using this framework, they extracted $f$-mode frequencies and determined the onset of the CFS instability for stars rotating up to the mass-shedding limit~\cite{PhysRevD.102.064026,PhysRevLett.125.111106}. More recently, Gao and Shibata calculated the complex quasi-normal-mode spectrum of rapidly rotating relativistic stars using a spectral method, including fluid modes and strongly damped $w$-modes \cite{Gao2026}. The existence of $w$-mode instability in rotating ultracompact stars does not by itself establish whether it limits the energy available to power sGRBs. This requires assessing the mode stability along the energy-extraction paths of post-merger ergostars, whose initial configurations are constrained by the angular-momentum budget of BNS mergers \cite{Xia2026}.

In this work, we investigate the $w$-mode spectrum and instability of strangeon ergostars. We derive the linearized perturbation equations in the Hilbert gauge and develop a spectral solver to compute the complex eigenvalues of stellar oscillation modes. We calculate the complex eigenvalues of the axial-led and polar-led $w$-mode branches connected to the nonrotating $l=m=2$ modes along the maximum-energy-extraction paths of twelve strangeon EOS models in our previous study \cite{Xia2026}. We then examine additional sequences at the central density of each path endpoint to locate the onset of instability and compare the critical angular velocities with those of the reference configurations. These calculations allow us to assess whether the considered $w$-mode instabilities provide a competing channel for the rotational energy available to electromagnetic extraction.

The paper is organized as follows. In Section~\ref{sec: eos}, we introduce the strangeon EOS and the stellar models considered in this work. In Section~\ref{sec: formulation}, we outline the perturbation framework and numerical method. In Section~\ref{sec: results}, we present the $w$-mode spectra along the maximum-energy-extraction paths and examine the onset of instability at higher rotation rates. We summarize our main findings and discuss the implications of our results for rotational energy extraction in Section~\ref{sec: conclusions}. Further technical details and numerical validation are provided in Appendix~\ref{app: perturbation}.

\section{Strangeon Equation of State}
\label{sec: eos}

We adopt the phenomenological Lennard--Jones model of strangeon matter used in previous studies of the structure, deformation, and oscillations of SnSs \cite{Lai2009,Gao_2022,2022MNRAS.516.6172L,Xia2026}. The interaction between two strangeons is described by
\begin{equation}
    U(r)=4\epsilon\left[\left(\frac{\sigma}{r}\right)^{12}-\left(\frac{\sigma}{r}\right)^6\right],
    \label{eq:LJ_potential}
\end{equation}
where $\epsilon$ denotes the depth of the potential well and $\sigma$ is the characteristic separation at which the potential vanishes. Assuming a crystalline arrangement of strangeons, the zero-temperature energy density and pressure are
\begin{align}
    \rho &=2\epsilon\left(A_{12}\sigma^{12}n^5-A_6\sigma^6n^3\right)+nN_qm_q,
    \label{eq:eos_rho_original}\\
    P &=n^2\frac{\dd(\rho/n)}{\dd n}=4\epsilon\left(2A_{12}\sigma^{12}n^5-A_6\sigma^6n^3\right),
    \label{eq:eos_p_original}
\end{align}
where $n$ is the strangeon number density, $A_{12}=6.2$ and $A_6=8.4$ are lattice sums, $N_q$ is the number of constituent quarks in each strangeon, and $m_q$ is the constituent quark mass \cite{Lai2009,Gao_2022,2022MNRAS.516.6172L}. Following our previous work, we take $N_q=18$ and $m_q=300~{\rm MeV}$.

For the present calculation, we adopt the two-parameter reformulation introduced by Zhang \emph{et al.} in their study of hybrid SnSs \cite{Zhang_2023}. The microscopic length scale $\sigma$ can be replaced by the surface baryon number density,
\begin{equation}
    n_{\rm sur}=\left(\frac{A_6}{2A_{12}}\right)^{1/2}\frac{N_q}{3\sigma^3},
    \label{eq:nsur}
\end{equation}
which corresponds to the density at which the pressure vanishes. Introducing
\begin{equation}
    \tilde{\epsilon}\equiv\frac{\epsilon}{N_q},\qquad \bar n\equiv\frac{N_q n}{n_{\rm sur}},
\end{equation}
the strangeon EOS can be rewritten as
\begin{align}
    \frac{\rho}{n_{\rm sur}}&=\frac{a}{9}\tilde{\epsilon}\left(\frac{\bar n^5}{18}-\bar n^3\right)+m_q\bar n,
    \label{eq:eos_rho}\\
    \frac{P}{n_{\rm sur}}&=\frac{2a}{9}\tilde{\epsilon}\left(\frac{\bar n^5}{9}-\bar n^3\right),
    \label{eq:eos_pressure}
\end{align}
where
\begin{equation}
    a=\frac{A_6^2}{A_{12}}\simeq11.38.
\end{equation}
The pressure vanishes at $\bar n=3$. In this form, the EOS is specified by the two parameters $n_{\rm sur}$ and $\tilde{\epsilon}$, with the explicit dependence on the assumed number of quarks in each strangeon absorbed into the redefinition \cite{Zhang_2023}. The parameter $\tilde{\epsilon}$ characterizes the interaction strength and primarily controls the stiffness of the EOS, whereas $n_{\rm sur}$ sets the density scale of the self-bound matter. It is not surprising that this ultrabaric strangeon matter could be superluminal, given the non-relativistic potential description of the interactions involved, and the interpretation of the adiabatic sound speed in this phenomenological EOS has been discussed in detail in our previous work \cite{Xia2026, Qi2025}.

\begin{figure}[t]
    \centering
    \includegraphics[width=\columnwidth]{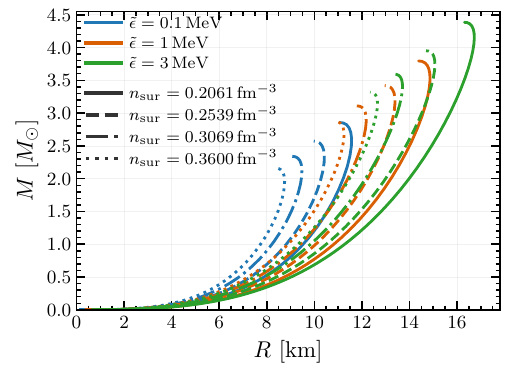}
    \caption{Mass--radius relations of nonrotating SnSs for the twelve EOS models listed in Table~\ref{tab:models}. Colors and line styles distinguish different values of $\tilde{\epsilon}$ and $n_{\rm sur}$, respectively, as indicated in the legends. Each sequence is shown up to its maximum-mass configuration.}
    \label{fig:mr}
\end{figure}

In this work, we consider twelve representative EOS models, labeled A--L, used to evaluate rotational energy extraction in our previous study \cite{Xia2026}. For each of the three values $\tilde{\epsilon}=0.1$, $1.0$, and $3.0~{\rm MeV}$, we consider four values of the surface baryon number density: $n_{\rm sur}=0.2061$, $0.2539$, $0.3069$, and $0.36~{\rm fm}^{-3}$. The corresponding model labels and EOS parameters are listed in Table~\ref{tab:models}. The mass--radius relations of the corresponding nonrotating SnS sequences are shown in Fig.~\ref{fig:mr}.

\begin{table}[t]
    \centering
    \small
    \setlength{\tabcolsep}{2pt}
    \caption{Parameters of the twelve strangeon EOS models
    adopted in this work \cite{Xia2026}.}
    \label{tab:models}
    \begin{tabular}{ccc}
        \hline\hline
        Model & $\tilde{\epsilon}$ (MeV)
              & $n_{\rm sur}$ (${\rm fm}^{-3}$) \\
        \hline
        A & 0.1 & 0.2061 \\
        B & 0.1 & 0.2539 \\
        C & 0.1 & 0.3069 \\
        D & 0.1 & 0.3600 \\
        \hline
        E & 1.0 & 0.2061 \\
        F & 1.0 & 0.2539 \\
        G & 1.0 & 0.3069 \\
        H & 1.0 & 0.3600 \\
        \hline
        I & 3.0 & 0.2061 \\
        J & 3.0 & 0.2539 \\
        K & 3.0 & 0.3069 \\
        L & 3.0 & 0.3600 \\
        \hline\hline
    \end{tabular}
\end{table}

\section{Formulation}
\label{sec: formulation}

The equilibrium configurations of uniformly rotating SnSs are constructed using LORENE \cite{Bonazzola1998}. Their metric is written as
\begin{equation}
\dd s^2=-e^{\gamma+\beta}\dd t^2+e^{2\alpha}(\dd r^2+r^2\dd \theta^2)+e^{\gamma-\beta}r^2\sin^2\theta(\dd \phi-\omega \dd t)^2,
\end{equation}
where the metric potentials depend on $r$ and $\theta$. The stellar fluid is described by a perfect-fluid energy-momentum tensor with four-velocity $u^a=u^t(1,0,0,\Omega)$. Throughout this work, we adopt geometrized units $G=c=1$.

To compute the quasi-normal modes of rapidly rotating stars, we solve the coupled spacetime and fluid perturbation equations in the Hilbert gauge, $\nabla^a\psi_{ab}=0$, where $\psi_{ab}=h_{ab}-\frac12g_{ab}h$ is the trace-reversed metric perturbation. We assume harmonic dependence $e^{i(\sigma t+m\phi)}$, with complex eigenvalue $\sigma$. Under this convention, $\mathrm{Im}(\sigma)>0$ corresponds to damping, while $\mathrm{Im}(\sigma)<0$ indicates instability.

The perturbation variables are expanded in tensor spherical harmonics, and their radial dependence is discretized using Chebyshev spectral methods. The resulting equations take the form of a quadratic matrix eigenvalue problem,
\begin{equation}
(\mathbf A_0+\sigma \mathbf A_1+\sigma^2 \mathbf A_2)\boldsymbol{x}=0.
\end{equation}
Regularity is imposed at the stellar center and rotation axis, while the finite surface density of strangeon matter is accounted for in the stellar-surface matching conditions. At spatial infinity, the purely outgoing-wave condition is imposed using exterior complex scaling (ECS) \cite{Andersson1995}. The axial-led and polar-led $w$-mode branches are identified by continuously tracing the corresponding fundamental $l=m=2$ modes from the nonrotating limit. The perturbation equations, spectral construction, and boundary conditions are detailed in Appendix~\ref{app: perturbation}.

\section{Results}
\label{sec: results}

\subsection{$w$-mode stability along maximum-energy-extraction paths}
\label{sec: path}

We first examine the $w$-mode spectrum along the maximum-energy-extraction paths of the twelve EOS models identified in our previous study \cite{Xia2026}. For each EOS, the selected path maximizes the decrease in gravitational mass among the allowed constant-baryon-mass sequences. Along these paths, the central density increases as the star loses angular momentum. For each equilibrium configuration along these paths, we calculate the complex eigenvalues of the axial-led and polar-led $w$-mode branches. We denote the central densities at the initial and final configurations by $\rho_{c,0}$ and $\rho_{c,\mathrm{end}}$, respectively. Since these endpoint densities differ among EOS models, we parameterize each path by its normalized logarithmic central density,
\begin{equation}
    \lambda=\frac{\ln(\rho_c/\rho_{c,0})}{\ln(\rho_{c,\mathrm{end}}/\rho_{c,0})},
    \qquad 0\leq\lambda\leq1,
    \label{eq:path_parameter}
\end{equation}
where $\rho_c=\varepsilon_c/c^2$, with $\varepsilon_c$ the central energy density. The reference densities $\rho_{c,0}$ and $\rho_{c,\mathrm{end}}$ are those at the low- and high-density endpoints of the original extraction path, corresponding to $\lambda=0$ and $\lambda=1$, respectively.

\begin{figure*}[t]
    \centering
    \includegraphics[width=\textwidth]{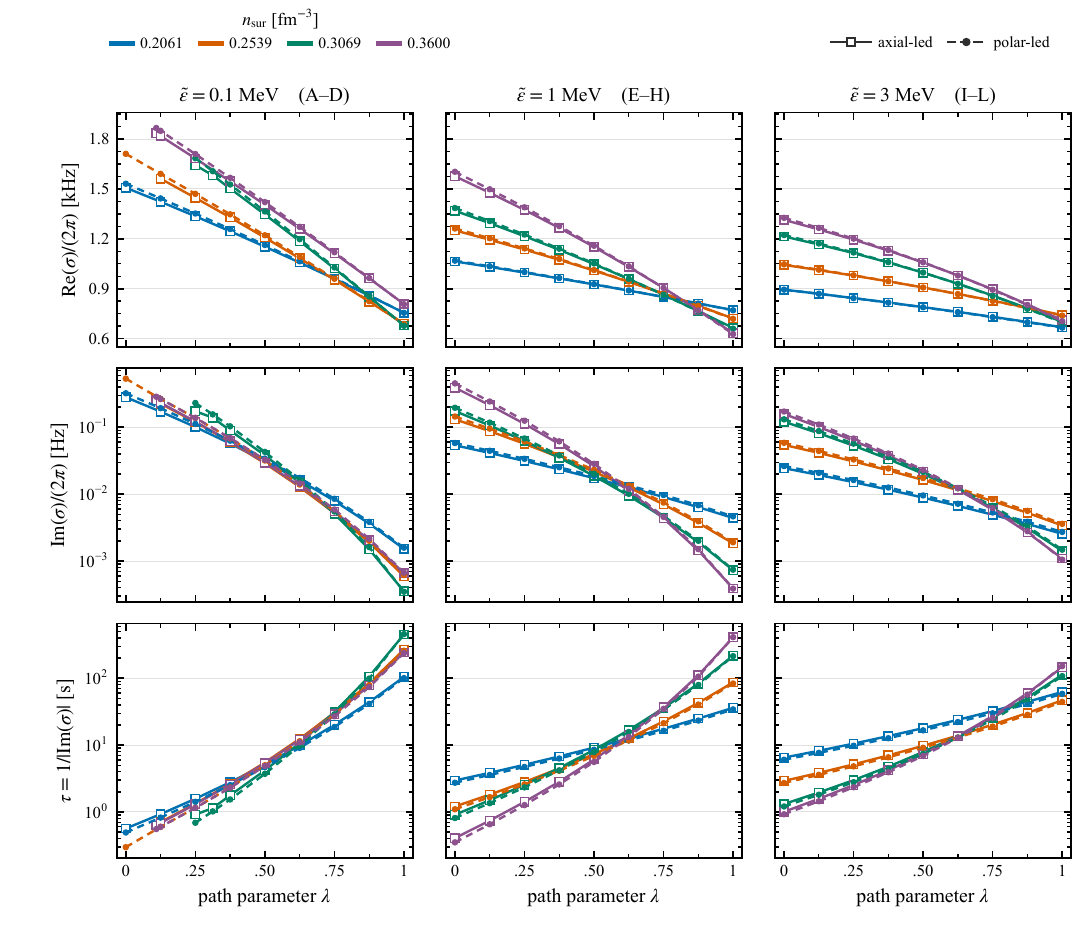}
    \caption{Complex eigenvalues and amplitude damping times of the axial-led and polar-led $w$-mode branches along the maximum-energy-extraction paths of the twelve strangeon EOS models. The columns correspond to $\tilde{\epsilon}=0.1$, $1.0$, and $3.0~{\rm MeV}$, respectively. From top to bottom, the rows show $\mathrm{Re}(\sigma)/(2\pi)$ in kHz, $\mathrm{Im}(\sigma)/(2\pi)$ in Hz, and $\tau=1/|\mathrm{Im}(\sigma)|$ in seconds. Colors distinguish the four values of $n_{\rm sur}$. Solid curves with open squares and dashed curves with filled circles denote the axial-led and polar-led branches, respectively. The path coordinate $\lambda$ is the normalized logarithmic central density defined in Eq.~\eqref{eq:path_parameter}.}
    \label{fig:energy_path}
\end{figure*}

Figure~\ref{fig:energy_path} shows the real and imaginary parts of the complex eigenvalues divided by $2\pi$, together with the amplitude damping times $\tau=1/|\mathrm{Im}(\sigma)|$. The oscillation frequencies generally decrease with increasing $\lambda$, accompanied by a decrease in the damping rates and a corresponding increase in the damping times. The axial-led and polar-led branches exhibit similar behavior, with their oscillation frequencies remaining close throughout the paths. Most importantly, $\mathrm{Im}(\sigma)$ remains positive for both branches in all twelve models, indicating that neither of the examined $w$-mode branches becomes unstable along these paths.

\begin{table}[t]
    \centering
    \small
    \setlength{\tabcolsep}{4pt}
    \caption{Frequencies and damping times of the axial-led and polar-led $w$-mode branches at the highest-central-density configuration ($\lambda=1$) along the maximum-energy-extraction path of each EOS model. Here $f=\mathrm{Re}(\sigma)/(2\pi)$ and $\tau=1/|\mathrm{Im}(\sigma)|$.}
    \label{tab:energy_path}
    \begin{tabular}{c rr rr}
        \hline\hline
        & \multicolumn{2}{c}{$f$ (Hz)} & \multicolumn{2}{c}{$\tau$ (s)} \\
        Model & axial-led & polar-led & axial-led & polar-led \\
        \hline
        A & 755.0 & 754.9 & 104.7 & 100.5 \\
        B & 689.0 & 687.6 & 267.9 & 257.6 \\
        C & 678.2 & 676.3 & 451.9 & 457.0 \\
        D & 807.7 & 806.3 & 242.4 & 236.8 \\
        \hline
        E & 771.2 & 770.0 & 36.1 & 34.2 \\
        F & 721.1 & 718.6 & 86.6 & 83.3 \\
        G & 663.8 & 659.6 & 217.6 & 214.1 \\
        H & 632.6 & 627.2 & 408.3 & 410.7 \\
        \hline
        I & 668.2 & 665.7 & 62.4 & 58.4 \\
        J & 740.8 & 738.2 & 46.8 & 44.3 \\
        K & 700.5 & 696.4 & 112.4 & 107.5 \\
        L & 709.9 & 704.5 & 145.8 & 151.7 \\
        \hline\hline
    \end{tabular}
\end{table}

Table~\ref{tab:energy_path} summarizes the oscillation frequencies and damping times at the highest-central-density configuration ($\lambda=1$) of each path. The frequencies range from approximately $0.63$ to $0.81~{\rm kHz}$, while the damping times range from about $34$ to $457~{\rm s}$. Although the damping rates decrease toward the high-density end of the paths, neither branch reaches the neutral stability condition. To compare these weakly damped configurations with the onset of instability, we next vary the angular velocity while holding the central density of each path endpoint fixed.

\subsection{Critical rotation rates for $w$-mode instability}
\label{sec: rates}

We next examine the onset of $w$-mode instability along sequences of fixed central density. For each EOS model, we select the highest-central-density configuration at $\lambda=1$ on its maximum-energy-extraction path, where the two branches have the weakest damping along that path. We denote its central density and angular velocity by $\rho_{c,1}$ and $\Omega_1$, respectively. Keeping $\rho_c=\rho_{c,1}$ fixed, we calculate the complex eigenvalues as functions of the stellar angular velocity up to $\Omega_{\max}\simeq0.95\Omega_{\rm K}$, where $\Omega_{\rm K}$ is the mass-shedding angular velocity of the corresponding fixed-central-density sequence. This upper limit is adopted to maintain surface fitting in the perturbation calculation, as discussed in Appendix~\ref{app: boundary_conditions}. For branches that become unstable within this range, we identify the critical angular velocity $\Omega_{\rm crit}$ at which $\mathrm{Im}(\sigma)$ changes from positive to negative. The relative position of the path endpoint with respect to this threshold is characterized by
\begin{equation}
\eta_\Omega=\frac{\Omega_1}{\Omega_{\rm crit}}.
\label{eq:critical_rotation_ratio}
\end{equation}

\begin{table*}[t]
    \centering
    \caption{Critical angular velocities of the axial-led and polar-led $w$-mode branches along sequences with central density fixed at $\rho_{c,1}$. The angular velocity of the corresponding path endpoint is denoted by $\Omega_1$, and $\eta_\Omega=\Omega_1/\Omega_{\rm crit}$. All angular velocities are given in ${\rm rad\,s^{-1}}$. A dash indicates that no instability onset was identified up to $\Omega_{\max}\simeq0.95\Omega_{\rm K}$.}
    \label{tab:critical_rotation}
    \begin{ruledtabular}
    \begin{tabular}{c r rr rr}
        & & \multicolumn{2}{c}{Axial-led} & \multicolumn{2}{c}{Polar-led} \\
        Model & $\Omega_1$ & $\Omega_{\rm crit}$ & $\eta_\Omega$ (\%) & $\Omega_{\rm crit}$ & $\eta_\Omega$ (\%) \\
        \hline
        A & 11839.9 & --- & --- & --- & --- \\
        B & 13438.1 & --- & --- & --- & --- \\
        C & 14883.1 & --- & --- & --- & --- \\
        D & 16027.3 & --- & --- & --- & --- \\
        \hline
        E & 8222.9 & 9525.5 & 86.32 & 9510.6 & 86.46 \\
        F & 9454.4 & 10618.0 & 89.04 & 10604.2 & 89.16 \\
        G & 10669.1 & 11704.6 & 91.15 & 11686.6 & 91.29 \\
        H & 11742.4 & 12706.4 & 92.41 & 12686.1 & 92.56 \\
        \hline
        I & 6760.9 & 8008.3 & 84.42 & 8023.8 & 84.26 \\
        J & 7716.8 & 9036.4 & 85.40 & 9048.1 & 85.29 \\
        K & 8657.9 & 9896.0 & 87.49 & 9910.2 & 87.36 \\
        L & 9557.0 & 10812.6 & 88.39 & 10792.1 & 88.56 \\
    \end{tabular}
    \end{ruledtabular}
\end{table*}

Table~\ref{tab:critical_rotation} summarizes the results for all twelve EOS models. Both the axial-led and polar-led branches become unstable at sufficiently high rotation rates for Models E--L, whereas neither branch becomes unstable for Models A--D up to the scan limit $\Omega_{\max}$. For every branch with an identified threshold, $\Omega_{\rm crit}$ exceeds $\Omega_1$, with $\eta_\Omega$ ranging from approximately $0.843$ to $0.926$. Within each of the $\tilde{\epsilon}=1$ and $3~{\rm MeV}$ families, this ratio increases with $n_{\rm sur}$. The critical angular velocities of the axial-led and polar-led branches are close, differing by less than $0.2\%$ for each model.

\begin{figure*}[t]
    \centering
    \includegraphics[width=\textwidth]{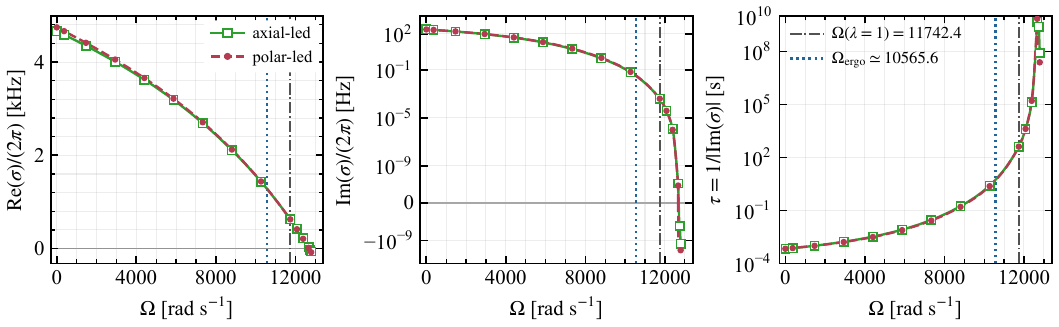}
    \caption{Complex eigenvalues and amplitude damping or growth times of the axial-led and polar-led $w$-mode branches for Model H along a sequence with central density fixed at that of the extraction-path endpoint ($\lambda=1$). From left to right, the panels show $\mathrm{Re}(\sigma)/(2\pi)$ in kHz, $\mathrm{Im}(\sigma)/(2\pi)$ in Hz, and $\tau=1/|\mathrm{Im}(\sigma)|$ in seconds, as functions of the angular velocity $\Omega$ in ${\rm rad\,s^{-1}}$. Solid curves with open squares and dashed curves with filled circles denote the axial-led and polar-led branches, respectively. Positive and negative imaginary parts correspond to damping and growth, respectively. The vertical gray dash-dotted line marks the angular velocity of the extraction-path endpoint ($\lambda=1$), $\Omega_1=11742.4~{\rm rad\,s^{-1}}$. The blue dotted line marks $\Omega_{\rm ergo}\simeq10565.6~{\rm rad\,s^{-1}}$, at which an ergoregion first appears along this fixed-central-density sequence.}
    \label{fig: H_rhoend}
\end{figure*}

\begin{figure*}[t]
    \centering
    \includegraphics[width=\textwidth]{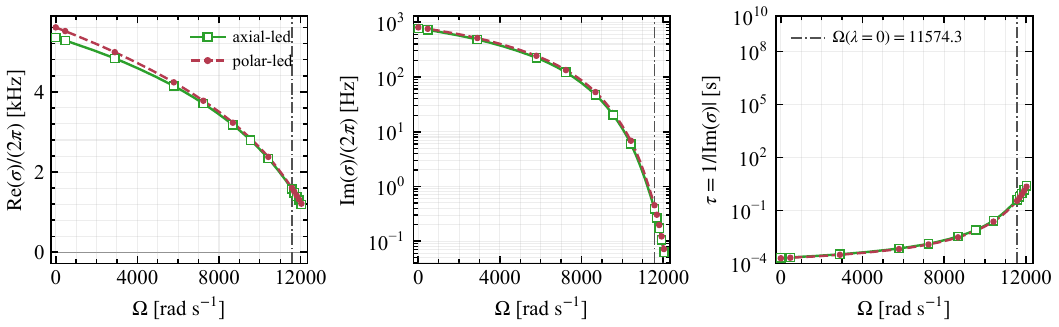}
    \caption{Same as Fig.~\ref{fig: H_rhoend}, but for the fixed-central-density sequence passing through the low-density endpoint ($\lambda=0$) of Model H's maximum-energy-extraction path. The vertical gray dash-dotted line marks $\Omega(\lambda=0)=11574.3~{\rm rad\,s^{-1}}$, which coincides with the onset of ergoregion formation along this sequence.}
    \label{fig: H_rho0}
\end{figure*}

Model H has the largest value of $\eta_\Omega$ among the models with identified instability thresholds and is therefore chosen as a representative case in Fig.~\ref{fig: H_rhoend}. Along its fixed-central-density sequence, the real parts of both eigenvalues decrease with increasing angular velocity, while their imaginary parts approach zero from above. At the high-rotation end, both the real and imaginary parts of the eigenvalues pass through zero, marking the transition to growing modes. The amplitude timescale $\tau=1/|\mathrm{Im}(\sigma)|$ increases sharply near neutrality and represents a damping time or a growth time on the corresponding side of the transition.

For Model H, an ergoregion first appears along the fixed-central-density sequence at $\Omega_{\rm ergo}\simeq10565.6~{\rm rad\,s^{-1}}$, marked by the blue dotted line in Fig.~\ref{fig: H_rhoend}. The critical angular velocities are $\Omega_{\rm crit}^{\rm a}=12706.4~{\rm rad\,s^{-1}}$ and $\Omega_{\rm crit}^{\rm p}=12686.1~{\rm rad\,s^{-1}}$ for the axial-led and polar-led branches, respectively. The extraction-path endpoint, marked by the gray dash-dotted lines, has $\Omega_1=11742.4~{\rm rad\,s^{-1}}$, corresponding to $\eta_\Omega^{\rm a}=0.9241$ and $\eta_\Omega^{\rm p}=0.9256$. Thus, even for the closest case in this comparison, the endpoint remains below both instability thresholds.

For comparison, we also examine the fixed-central-density sequence passing through the low-density endpoint ($\lambda=0$) of Model H's extraction path, as shown in Fig.~\ref{fig: H_rho0}. At this central density, the endpoint angular velocity $\Omega(\lambda=0)=11574.3~{\rm rad\,s^{-1}}$ coincides with the onset of ergoregion formation. Unlike the $\lambda=1$ sequence, both $w$-mode branches remain damped throughout the rotation range examined. This comparison further illustrates that the formation of an ergoregion does not necessarily trigger instability in the particular $w$-mode branches considered here.

\section{Conclusions}
\label{sec: conclusions}

Our results provide further insight into the relation between the ergoregion and $w$-mode instability. The possibility of a gravitational-wave-driven $w$-mode instability in rotating ultracompact stars was demonstrated by Kokkotas \emph{et al.} \cite{Kokkotas_2004}, who applied a first-order slow-rotation formalism to uniform-density models at rotation rates extending toward the Kepler limit and found that axial $w$-modes can become unstable after the formation of an ergoregion. Building on their work, we investigate the stability of $w$-modes in rapidly rotating strangeon ergostars. The axial-led and polar-led $w$-mode branches examined here remain damped throughout the maximum-energy-extraction paths. For Models E--L, both branches become unstable along the additional fixed-central-density sequences, but only at angular velocities above those of the corresponding path endpoints.

The relevance of a $w$-mode instability to an ergostar central engine depends on whether it is encountered during the release of rotational energy. The unstable branches found on the additional fixed-central-density sequences show that the occurrence of this instability in the broader set of equilibrium configurations need not preclude the energy-extraction scenario. Along the maximum-energy-extraction paths themselves, the two branches remain damped, and the least-damped configurations lie below their corresponding instability thresholds wherever these thresholds are identified. Within the equilibrium-sequence description adopted here, these modes therefore do not develop the instability-driven growth that could redirect rotational energy into gravitational radiation. The rotational-energy reservoirs estimated in our previous study \cite{Xia2026} are consequently not further restricted by this particular mechanism.

The EOS dependence of the stability margin can be examined through the ratio $\eta_\Omega=\Omega_1/\Omega_{\rm crit}$. Among Models E--L, $\eta_\Omega$ increases with $n_{\rm sur}$ within each interaction-strength family, while the $\tilde{\epsilon}=3~{\rm MeV}$ models have smaller ratios than the $\tilde{\epsilon}=1~{\rm MeV}$ models at the same $n_{\rm sur}$. These trends reflect both the mode spectrum and the location of the reference configuration selected by the energy-extraction path. Even for Model H, which gives the largest ratio, the lower of the two critical angular velocities remains approximately $8\%$ above $\Omega_1$.

Beyond the stability of individual oscillation modes, the astrophysical implications of ergostars also depend on how their rotational energy is released into the surrounding environment. While the $w$-mode spectrum characterizes the gravitational response of the star, electromagnetic energy extraction is closely related to the structure and evolution of its magnetosphere \cite{Komissarov2004,Komissarov2005,Tsokaros2020c}. As discussed in our previous work \cite{Xia2026}, the self-bound surface of strangeon matter may also play an important role in suppressing baryon contamination of relativistic outflows. Further investigations of the interplay between gravitational-wave emission and electromagnetic processes would help establish a more comprehensive picture of strangeon ergostars as potential central engines of sGRBs.

\begin{acknowledgments}
H.~X. thanks Yong Gao for helpful discussions and suggestions during the development of the numerical method presented in this work. This work was supported by the National Natural Science Foundation of China (Grant No. 8200910057, 12447148), and the National SKA Program of China (2020SKA0120100). E. Z. is supported by NSFC Grant No. 12203017 and the National SKA Program of China No. 2020SKA0120300.
\end{acknowledgments}

\appendix

\section{Perturbation equations and numerical implementation}
\label{app: perturbation}

\subsection{Linearized Einstein--fluid equations}
\label{app: equations}

We consider linear perturbations of the uniformly rotating equilibrium configurations introduced in Section~\ref{sec: formulation}. Denoting the metric perturbation by $h_{ab}=\delta g_{ab}$, we introduce its trace-reversed form
\begin{equation}
\psi_{ab}=h_{ab}-\frac12g_{ab}h,\qquad h=g^{ab}h_{ab}.
\end{equation}
We adopt the Hilbert gauge, which is also employed by Kr\"uger and Kokkotas in their time-domain study of rapidly rotating relativistic stars \cite{PhysRevD.102.064026},
\begin{equation}
C_b\equiv\nabla^a\psi_{ab}=0.
\label{eq:hilbert}
\end{equation}
The linearized Einstein equations can then be expressed in terms of the operator
\begin{align}
(\mathcal D\psi)_{ab}={}&\Box\psi_{ab}+2R_{acbd}\psi^{cd}
-R_a{}^c\psi_{cb}-R_b{}^c\psi_{ca}\nonumber\\
&+R\psi_{ab}-g_{ab}R^{cd}\psi_{cd},
\label{eq:einstein_operator}
\end{align}
where $\Box=g^{cd}\nabla_c\nabla_d$. In particular,
\begin{equation}
\delta G_{ab}=-\frac12(\mathcal D\psi)_{ab}+\nabla_{(a}C_{b)}-\frac12g_{ab}\nabla_c C^c.
\end{equation}
The field equations in the Hilbert gauge therefore reduce to
\begin{equation}
(\mathcal D\psi)_{ab}+16\pi\delta T_{ab}=0.
\label{eq:einstein}
\end{equation}

For the fluid perturbations, we define
\begin{equation}
w=\rho+P,\qquad q_h=h_{ab}u^a u^b,\qquad H_1=\frac{\delta P}{w}.
\end{equation}
We then introduce a velocity perturbation $v_a$ orthogonal to the background four-velocity, $u^av_a=0$, through
\begin{equation}
\delta u_a=v_a-\frac12q_hu_a.
\end{equation}
This parametrization automatically satisfies the linearized normalization condition $\delta(u^au_a)=0$. For a barotropic EOS, with $c_s^2=dP/d\rho$, the perturbation of the energy-momentum tensor becomes
\begin{align}
\delta T_{ab}={}&Ph_{ab}-wq_hu_au_b+w(v_au_b+u_av_b)\nonumber\\
&+wH_1\left[g_{ab}+\left(1+c_s^{-2}\right)u_au_b\right].
\end{align}
Since $u^av_a=u^t(v_t+\Omega v_\phi)=0$, the independent fluid perturbation variables can be chosen as
\begin{equation}
\boldsymbol z=(H_1,v_r,v_\theta,v_\phi).
\end{equation}

The fluid perturbations additionally satisfy the linearized energy-momentum conservation equations. Defining the mixed-index perturbation
\begin{equation}
\tau^a{}_b\equiv\delta T^a{}_b=g^{ac}\delta T_{cb}-h^{ac}T_{cb},
\end{equation}
we obtain
\begin{equation}
\delta(\nabla_aT^a{}_b)=\nabla_a\tau^a{}_b+\delta\Gamma^a{}_{ac}T^c{}_b-\delta\Gamma^c{}_{ab}T^a{}_c=0,
\label{eq:fluid_conservation}
\end{equation}
where
\begin{equation}
\delta\Gamma^c{}_{ab}=\frac12g^{cd}\left(\nabla_ah_{bd}+\nabla_bh_{ad}-\nabla_dh_{ab}\right).
\end{equation}
In the stellar interior, we solve the ten reduced Einstein equations \eqref{eq:einstein} together with the four fluid conservation equations \eqref{eq:fluid_conservation}. In the vacuum exterior, only the metric equations remain. For each computed mode, we evaluate the Hilbert-gauge residual $C_b$ and the residual of the full linearized Einstein equations.

\subsection{Tensor-harmonic expansion and spectral discretization}
\label{app: spectral_method}

We expand the perturbation variables in tensor spherical harmonics and discretize their radial dependence using a spectral method. For numerical convenience, the tensor harmonics are constructed by coupling scalar spherical harmonics with irreducible tensor bases through Clebsch--Gordan coefficients \cite{Thorne1980}.

First, we introduce the following basis one-forms
\begin{equation}
\begin{aligned}
e_t &= \dd t,\qquad e_0=\dd z,\\
e_{\pm1}
&=\mp\frac{\dd x\pm i\,\dd y}{\sqrt{2}}\\
&=\mp e^{\pm i\phi}
\frac{\dd\varpi\pm i\varpi\,\dd\phi}{\sqrt{2}},
\end{aligned}
\end{equation}
where $\varpi=r\sin\theta$. Under spatial rotations, $e_t$ transforms as a scalar ($S=0$), while the three spatial one-forms $e_s$ form a vector representation ($S=1$). The representation carried by a spacetime covector is therefore $\mathbf{1}\oplus\mathbf{3}$.

Since the metric perturbation is symmetric, its tensor representation decomposes as
\begin{equation}
\operatorname{Sym}^2(\mathbf{1}\oplus\mathbf{3})
=\mathbf{1}\oplus\mathbf{1}\oplus\mathbf{3}\oplus\mathbf{5}.
\end{equation}
The labels $\mathrm{tt}$, $\mathrm{tv}$, and $\mathrm{sp}$ denote the temporal--temporal, temporal--spatial, and purely spatial tensor sectors, respectively. We distinguish the four irreducible families by $\alpha\in\{\mathrm{tt},\mathrm{tv},(\mathrm{sp},0),(\mathrm{sp},2)\}$, with respective spins $S_\alpha=0,1,0,2$. Their basis tensors are
\begin{align}
T^{(\mathrm{tt})}_0 &= e_t\otimes e_t,\\
T^{(\mathrm{tv})}_s &= \frac{1}{\sqrt{2}}\left(e_t\otimes e_s+e_s\otimes e_t\right),\\
T^{(\mathrm{sp},S)}_s &= \sum_{s_1,s_2}\langle 1s_1,1s_2|Ss\rangle e_{s_1}\otimes e_{s_2},
\end{align}
where $S=0,2$ in the last line correspond to the spatial trace and symmetric trace-free parts, respectively. Coupling these bases to scalar spherical harmonics through Clebsch--Gordan coefficients gives
\begin{equation}
\mathcal{Y}^{(\alpha)}_{JLm}
=\sum_{k,s}
\langle Lk,S_\alpha s|Jm\rangle
Y_{Lk}\,T^{(\alpha)}_s.
\end{equation}
The trace-reversed metric perturbation is then expanded as
\begin{equation}
\psi_{ab}
=e^{i\sigma t}
\sum_{\alpha,J,L}
f^{(\alpha)}_{JL}(r)
\bigl(\mathcal{Y}^{(\alpha)}_{JLm}\bigr)_{ab},
\end{equation}
where the sum includes the tensor sectors appropriate to each component. The scalar fluid variable $H_1$ and the spatial velocity perturbation are expanded in scalar and vector harmonics, respectively. In cylindrical components, all terms carry the common factor $e^{im\phi}$. For a tensor component with spin projection $s$, the remaining polar-angle dependence is proportional to $P_L^{m-s}(\cos\theta)$.

Axisymmetry and equatorial reflection symmetry allow the perturbations to be separated into independent sectors. For $m=2$, we refer to the sectors containing the nonrotating $l=2$ axial and polar modes as axial-led and polar-led, respectively. The corresponding angular-channel selection rules are summarized in Table~\ref{tab:angular_channels}.

\begin{table}[t]
    \centering
    \small
    \setlength{\tabcolsep}{4pt}
    \caption{Angular-channel selection rules for the $m=2$ axial-led and polar-led sectors.}
    \label{tab:angular_channels}
    \begin{tabular}{lccc}
        \hline\hline
        $S$ & Allowed $L$ & Axial-led & Polar-led \\
        \hline
        $0$ & $L=J$ & $J$ odd & $J$ even \\
        $1$ & $|J-1|\leq L\leq J+1$ & $L$ even & $L$ odd \\
        $2$ & $|J-2|\leq L\leq J+2$ & $L$ odd & $L$ even \\
        \hline\hline
    \end{tabular}
\end{table}

Rather than deriving and implementing the component equations individually, we construct the differential operators directly from their covariant definitions. As an example, the ten independent components of $\psi_{ab}$ can be collected into a vector $\boldsymbol\psi$. The action of the connection on this vector is represented by matrices $B_a$, defined through
\begin{equation}
(B_a\psi)_{bc}=-\Gamma^d{}_{ab}\psi_{dc}-\Gamma^d{}_{ac}\psi_{bd}.
\end{equation}
The covariant derivative then takes the matrix form
\begin{equation}
\nabla_a\boldsymbol\psi=(\partial_a+B_a)\boldsymbol\psi,
\end{equation}
and the covariant wave operator becomes
\begin{align}
\Box\boldsymbol\psi=g^{ab}\big[&
\partial_a\partial_b+B_a\partial_b+B_b\partial_a
+\partial_a B_b+B_aB_b\nonumber\\
&-\Gamma^c{}_{ab}(\partial_c+B_c)\big]\boldsymbol\psi.
\end{align}
The curvature terms in Eq.~\eqref{eq:einstein_operator} are likewise represented by matrices acting algebraically on $\boldsymbol\psi$. The coefficients of these operators are evaluated from the numerical background metric and its derivatives.

The resulting operators are projected onto the tensor-harmonic basis, while the radial dependence is discretized using a Chebyshev pseudospectral method. After angular truncation and radial discretization, all spatial differential operators are replaced by finite matrices. The temporal dependence introduces terms at most quadratic in $\sigma$, yielding a quadratic matrix eigenvalue problem of the form
\begin{equation}
\left(\mathbf A_0+\sigma\mathbf A_1+\sigma^2\mathbf A_2\right)\mathbf x=0,
\end{equation}
where $\boldsymbol x$ collects the discretized metric and fluid perturbation variables. The matrices $\mathbf A_0$, $\mathbf A_1$, and $\mathbf A_2$ contain the background-dependent coefficients, angular couplings, radial differentiation operators, and boundary conditions.

\subsection{Boundary conditions}
\label{app: boundary_conditions}

The perturbation equations are supplemented by regularity conditions at the stellar center and rotation axis, matching conditions at the stellar surface, and a purely outgoing-wave condition at spatial infinity.

\paragraph{Center and axis regularity.}

For a smooth field in a regular Cartesian tensor basis, a channel with orbital angular momentum $L$ has radial dependence $f_L(r)=O(r^L)$ near the center. In the exterior-complex-scaling calculations, we impose the corresponding leading-order condition
\begin{equation}
\left(\partial_r-\frac{L}{r}\right)f_L=0
\end{equation} 
at the innermost radial collocation point for the metric and enthalpy channels. For the velocity variables, the fluid conservation equations are retained at this point. Regularity on the rotation axis is enforced consistently with the angular-harmonic basis described in Appendix~\ref{app: spectral_method}.

\paragraph{Stellar-surface matching.}

The surface of a SnS requires particular attention because the energy density remains finite at zero pressure. Let the unperturbed stellar surface $\Sigma$ be described by
\begin{equation}
F(r,\theta)=r-S(\theta)=0.
\end{equation}
Denoting the outward unit normal by $n_a$, we write
\begin{equation}
n_a=\frac{\nabla_a F}{W_\Sigma},\qquad
W_\Sigma=\sqrt{g^{rr}+g^{\theta\theta}[S'(\theta)]^2}.
\end{equation}

Let $d$ be the signed proper distance from the surface, positive in the exterior, and let $\xi$ denote the normal displacement of the perturbed surface. Denoting the energy density at zero pressure by $\rho_{\rm sur}$, the discontinuous contribution to the background energy density near $\Sigma$ is proportional to $\rho_{\rm sur}\Theta(-d)$. Under the displacement $d\rightarrow d-\xi$, the corresponding first-order variation is
\begin{equation}
\rho_{\rm sur}\big[\Theta(\xi-d)-\Theta(-d)\big]
=\rho_{\rm sur}\xi\delta(d)+O(\xi^2).
\end{equation}
Since the background pressure vanishes at the surface, the singular contribution to the Eulerian energy-momentum perturbation is
\begin{equation}
\left.\delta T_{ab}\right|_{\rm surf}
=\rho_{\rm sur}u_au_b\,\xi\delta(d).
\end{equation}
The metric perturbation is continuous across the surface,
\begin{equation}
\left.\psi_{ab}^{\rm in}\right|_\Sigma
=\left.\psi_{ab}^{\rm out}\right|_\Sigma,
\label{eq:metric_continuity}
\end{equation}
while integrating the linearized Einstein equations across the displaced density discontinuity gives the derivative matching condition
\begin{equation}
\left.\left(\partial_r\psi_{ab}^{\rm in}-\partial_r\psi_{ab}^{\rm out}\right)\right|_\Sigma
=\frac{16\pi\rho_{\rm sur}\xi}{W_\Sigma}u_au_b.
\label{eq:surface_jump}
\end{equation}
Thus, unlike the case of an EOS whose density vanishes continuously at the surface, the normal displacement of a self-bound stellar surface contributes explicitly to the matching of the metric derivatives.

The fluid perturbations must also satisfy the free-surface condition that the Lagrangian pressure perturbation vanishes,
\begin{equation}
\Delta P=\delta P+\xi n^a\nabla_aP=0.
\end{equation}
Writing $P_n=n^a\nabla_aP$ and using $\delta P=wH_1$, the normal displacement is
\begin{equation}
\xi=-\frac{wH_1}{P_n}.
\end{equation}
The kinematic condition relating the fluid velocity to the motion of the surface is
\begin{equation}
n_a\delta u^a=i(\sigma+m\Omega)u^t\xi.
\end{equation}
Eliminating $\xi$ yields
\begin{equation}
P_n n_a\delta u^a+i(\sigma+m\Omega)u^t wH_1=0.
\label{eq:surface_fluid}
\end{equation}
Equations~\eqref{eq:metric_continuity}, \eqref{eq:surface_jump}, and \eqref{eq:surface_fluid} provide the surface conditions used to couple the interior fluid and metric perturbations to the exterior spacetime perturbations.

Applying these conditions requires the computational interface to coincide with the zero-enthalpy surface of the background star. Near the mass-shedding limit, the small equatorial radial enthalpy gradient makes the surface position sensitive to numerical errors. Freezing the angular shape of the LORENE domain mapping can facilitate continuation of the equilibrium sequence, but may leave an offset between the computational interface and the zero-enthalpy surface. For self-bound matter, this offset displaces the matching surface relative to the density discontinuity, even when the enthalpy change between successive equilibrium iterations is small. We therefore restrict the rotation scans to retain surface fitting for the perturbation calculation.

\begin{table*}[t]
    \centering
    \small
    \setlength{\tabcolsep}{5pt}
    \caption{Comparison of the $l=m=2$ axial and polar $w$-modes of nonrotating Model H at $\rho_c\simeq1.733\times10^{15}~{\rm g\,cm^{-3}}$, computed using independent static equations and our solver. The eigenvalue components $f_R$ and $f_I$ are given in Hz. The relative differences $\delta$ are measured with respect to the independent static solutions and expressed as percentages.}
    \label{tab:static_validation}
    \begin{tabular}{l c c c r}
        \hline\hline
        Mode & Component & Independent static solution & This work & $\delta$ (\%) \\
        \hline
        Axial & $f_R$ & 4663.51269262 & 4663.51269378 & $2.487\times10^{-8}$ \\
        Axial & $f_I$ & 235.665168804 & 235.665167251 & $6.590\times10^{-7}$ \\
        Polar & $f_R$ & 4744.35183021 & 4744.35154914 & $5.924\times10^{-6}$ \\
        Polar & $f_I$ & 232.985245353 & 232.985222031 & $1.001\times10^{-5}$ \\
        \hline\hline
    \end{tabular}
\end{table*}

\begin{table*}[t]
    \centering
    \small
    \setlength{\tabcolsep}{8pt}
    \caption{Convergence tests with respect to radial resolution and angular truncation order for the axial-led $w$-mode branch at the two rotating configurations of Model H.}
    \label{tab:mode_convergence}
    \begin{tabular}{c c r r r r}
        \hline\hline
        & & \multicolumn{2}{c}{$\Omega_{\rm low}$} & \multicolumn{2}{c}{$\Omega_{\rm near}$} \\
        $J_{\max}$ & $N_1/N_2/N_3$ & $f_R$ (Hz) & $f_I$ (Hz) & $f_R$ (Hz) & $f_I$ ($10^{-9}$ Hz) \\
        \hline
        23 & 57/73/113 & 4006.9900355614 & 95.09226657651 & $-63.975770891$ & $-1.936931314$ \\
        23 & 65/81/129 & 4006.9900341750 & 95.09226681063 & $-63.975768613$ & $-1.936941844$ \\
        23 & 65/89/145 & 4006.9900346149 & 95.09226597809 & $-63.975766589$ & $-1.936760426$ \\
        \hline
        23 & 45/57/89 & 4006.9900176076 & 95.09228868364 & $-63.975640504$ & $-1.930495824$ \\
        31 & 45/57/89 & 4006.9900176064 & 95.09228868339 & $-63.976877938$ & $-1.929167789$ \\
        33 & 45/57/89 & 4006.9900176080 & 95.09228868395 & $-63.976886147$ & $-1.929156403$ \\
        \hline\hline
    \end{tabular}
\end{table*}

\paragraph{Outgoing-wave condition.}

At spatial infinity, quasi-normal modes satisfy the purely outgoing-wave condition. With the time dependence $e^{i\sigma t}$ adopted here, the outgoing asymptotic solution behaves schematically as
\begin{equation}
\Psi_{\rm out}\sim e^{-i\sigma r_*},
\end{equation}
up to algebraic radial factors, where $r_*$ is an asymptotic wave coordinate. For damped modes with $\mathrm{Im}(\sigma)>0$, this solution grows exponentially along the real radial axis, making a direct boundary condition at large real radius numerically inconvenient.

We impose the outgoing condition using exterior complex scaling \cite{Andersson1995}. Beyond a radius $r_b$ outside the stellar surface, the radial coordinate is analytically continued along
\begin{equation}
r=r_b+s e^{i\vartheta},\qquad s\geq0,
\end{equation}
where $\vartheta$ is the complex-rotation angle. The angle is chosen such that
\begin{equation}
\mathrm{Im}\left(\sigma e^{i\vartheta}\right)<0,
\end{equation}
so that the outgoing solution decays exponentially along the rotated contour. This allows the outgoing-wave condition to be implemented as a decaying boundary condition at the outer end of the computational domain. The exterior spectral operators are evaluated along the complex contour, including the corresponding transformation of radial derivatives.

\subsection{Numerical validation}
\label{app: numerical_validation}

We first compare the nonrotating limit of our solver against independent solutions of the static perturbation equations for Model~H at a central density of $\rho_c\simeq1.733\times10^{15}~{\rm g\,cm^{-3}}$. The axial modes are obtained from the stellar axial perturbation equation \cite{Kokkotas1994}, which reduces to the Regge--Wheeler equation in the vacuum exterior \cite{RW1957}, while the polar modes are calculated using the Lindblom--Detweiler interior equations \cite{LindblomDetweiler1983,DetweilerLindblom1985} and the Zerilli equation in the exterior \cite{Zerilli1970}. Table~\ref{tab:static_validation} compares the real and imaginary eigenvalue components, denoted by $f_R=\mathrm{Re}(\sigma)/(2\pi)$ and $f_I=\mathrm{Im}(\sigma)/(2\pi)$. The largest relative difference is approximately $10^{-7}$, providing a validation of our solver in the nonrotating limit.

We then examine the dependence of the axial-led $w$-mode branch on the radial resolution and angular truncation order at two rotating configurations of Model~H with the same central density. The angular velocities are $\Omega_{\rm low}\simeq2935.6~{\rm rad\,s^{-1}}$ and $\Omega_{\rm near}\simeq12793.5~{\rm rad\,s^{-1}}$, with the latter lying close to the neutral point on its unstable side. Table~\ref{tab:mode_convergence} lists the eigenvalues for different angular truncations $J_{\max}$ and radial resolutions. The three node counts $N_1$, $N_2$, and $N_3$ refer to the stellar interior, the real exterior, and the ECS domain, respectively.

At fixed $J_{\max}=23$, increasing the radial resolution from $57/73/113$ to $65/81/129$ changes the near-neutral value of $f_I$ by approximately $5.4\times10^{-4}\%$. Angular convergence is evaluated on the common $45/57/89$ radial grid using $J_{\max}=23$, $31$, and $33$. Increasing $J_{\max}$ from 23 to 33 shifts $f_I$ by approximately $3.3\times10^{-10}\%$ at $\Omega_{\rm low}$ and by $0.069\%$ at $\Omega_{\rm near}$. The greater relative sensitivity of the near-neutral imaginary part to angular truncation motivates the use of higher angular resolutions in this regime.

\begin{table}[htb]
    \centering
    \small
    \setlength{\tabcolsep}{4pt}
    \caption{Reference and extended exterior configurations used in the ECS comparison. Here $R_{\rm eq}$ is the equatorial stellar radius and $L_{\rm bg}$ is the Legendre truncation order of the exterior background. The complex-ray lengths are expressed in units of $M_{\rm ref}$, the ADM mass of the nonrotating configuration on the same fixed-central-density sequence ($M_{\rm ref}\simeq2.862\,M_\odot$ for Model H).}
    \label{tab:ecs_validation}
    \begin{tabular}{l r r}
        \hline\hline
        Parameter & Reference & Extended \\
        \hline
        ECS starting radius & $12R_{\rm eq}$ & $20R_{\rm eq}$ \\
        $N_2/N_3$ & 81/129 & 89/145 \\
        $L_{\rm bg}$ & 8 & 16 \\
        Decay-length parameter $N_{\rm d}$ & 60 & 90 \\
        Ray length at $\Omega_{\rm low}$ & 239.094 & 358.641 \\
        Ray length at $\Omega_{\rm near}$ & 14975.171 & 22462.756 \\
        \hline\hline
    \end{tabular}
\end{table}

To test the exterior treatment, we shift the ECS starting radius from $12R_{\rm eq}$ to $20R_{\rm eq}$, increase the resolutions of both exterior domains as well as the Legendre truncation order of the exterior background, and extend the complex ray by a factor of $1.5$. The ECS ray length is specified by
\begin{equation}
    L_{\rm ray}=N_{\rm d}\ell_{\rm ref},
    \qquad
    \ell_{\rm ref}=\left[-\operatorname{Im}\!\left(\sigma_{\rm ref}e^{i\vartheta}\right)\right]^{-1},
    \label{eq:ecs_ray_length}
\end{equation}
where $\ell_{\rm ref}$ is the amplitude e-folding length of the asymptotic outgoing wave along the complex ray, evaluated using the input reference value $\sigma_{\rm ref}$. The parameters are listed in Table~\ref{tab:ecs_validation}. This combined refinement and domain extension alters the near-neutral value of $f_I$ by approximately $0.0094\%$.

Across all configurations in Table~\ref{tab:mode_convergence}, both eigenvalue components at $\Omega_{\rm low}$ vary by less than $2.4\times10^{-5}~{\rm Hz}$. At $\Omega_{\rm near}$, $f_I$ remains negative, with a relative spread of approximately $0.40\%$, supporting the identification of weak mode growth.

\bibliography{refs}

\end{document}